 \documentclass[preprint,floats,aps,epsfig,nofootinbib,amssymb]{revtex4}

\usepackage{slashed}
\usepackage{slashed}
\usepackage{graphicx,color}
\usepackage{epsfig}
\usepackage{subfigure}
\usepackage{epsfig}
\usepackage{dcolumn}
\usepackage{bm}
\usepackage{color}
\usepackage{multirow}
\usepackage{amsmath}

\newcommand{\beq}{\begin {equation}}
\newcommand{\eeq}{\end   {equation}}
\newcommand{\bea}{\begin {eqnarray}}
\newcommand{\eea}{\end   {eqnarray}}
\newcommand{\baa}{\begin {array}   }
\newcommand{\eaa}{\end   {array}   }
\newcommand{\bit}{\begin {itemize} }
\newcommand{\eit}{\end   {itemize} }
\newcommand{\be }{\begin {equation}}
\newcommand{\ee }{\end   {equation}}
\newcommand{\nn }{\nonumber        }

\newcommand{\VEV}[1]{\langle  #1 \rangle}

\begin{document}

\preprint{ACFI-T16-06}

\title{$SU(2) \times SU(2) \times U(1)$ Interpretation on the 750 GeV Diphoton Excess}

\author{Jing Ren\,\footnote{jren@physics.utoronto.ca}}
\author{Jiang-Hao Yu\,\footnote{jhyu@physics.umass.edu}}

\affiliation{$^*$ Department of Physics, University of Toronto, Toronto, Ontario, Canada M5S1A7}
\affiliation{$^\dagger$ Amherst Center for Fundamental Interactions, Department of Physics, University of Massachusetts-Amherst, Amherst, MA 01003, USA}



\begin{abstract}
	
We propose that the $SU(2) \times SU(2) \times U(1)$ ({\it a.k.a.} $G221$) models could provide us a 750 GeV scalar resonance that may account for the diphoton excess observed at the LHC while satisfying present collider constraints.
The neutral component of the $SU(2)_R$ scalar multiplet can be identified as the 750 GeV scalar.
In the lepto-phobic and fermio-phobic $G221$ models the new charged gauge boson $W'$ could be light, and we find that the diphoton decay width could be dominated by the loop contribution from the $W'$.
To initiate gluon fusion production, it is necessary to extend the $G221$ symmetry to the Pati-Salam and $SO(10)$ symmetry.
We investigate the possibilities that the light colored scalars or vectorlike fermions survive in the $SO(10)$ theory and provide large gluon fusion rate for the diphoton signature.
It is possible to test the $G221$ interpretation by direct searches of $W'$ using the multi-gauge boson production channel at the Run 2 LHC.

\end{abstract}

\maketitle


\section{Introduction}
\label{sec:intro}

Recently both ATLAS and CMS collaborations reported that a diphoton excess around 750 GeV has been found in pp collisions
at the LHC Run-2~\cite{ATLAS13, CMS13}. Under narrow width assumptions, ATLAS observed a diphoton excess at
$\,M_{\gamma\gamma}^{}=747$\,GeV with a $3.6\sigma$ local significance (3.2\,fb$^{-1}$ integrated luminosity)~\cite{ATLAS13} and CMS reported an excess
at $\,M_{\gamma\gamma}^{}=760$\,GeV with $2.6\sigma$ local significance (2.6\,fb$^{-1}$ integrated luminosity)~\cite{CMS13}. Assuming this decay channel includes just two photons but nothing else, the resonance can only be a spin-0 or spin-2 particle. Given null results of resonance searches in other channels at 13\,TeV (including diphoton at 8\,TeV), one favorable minimal interpretation of this diphoton excess is a (pseudo)scalar particle that couples to photons and gluons. Under narrow width assumptions, the cross section at 13TeV would be around~\cite{ATLAS13, CMS13}
\begin{eqnarray}
\sigma(pp\to S\to \gamma\gamma)\sim 5-10\,\textrm{fb}\,.
\end{eqnarray}
This signal strength requires a large diphoton width as well as a large gluon fusion production cross section, which cannot be achieved with the standard model (SM) particles and the (pseudo)scalar resonance but nothing else. If confirmed by future data, the excess may imply a whole new sector of new physics beyond SM and deserve exploration in different context of new physics models~\cite{Diphoton:all,Dey:2015bur,Dasgupta:2015pbr,Deppisch:2016scs,Berlin:2016hqw,Hati:2016thk,Aydemir:2016qqj}.

The $SU(2) \times SU(2) \times U(1)$ ({\it a.k.a.} $G221$) models~\cite{Hsieh:2010zr,Cao:2012ng} are the minimal extension of the SM gauge group that incorporates both the $W'$ and $Z'$ bosons in effective theory. Various models have been considered in this framework: left-right (LR)~\cite{Mohapatra:1974gc, Mohapatra:1974hk, Mohapatra:1980yp},
lepto-phobic (LP), hadro-phobic (HP), fermio-phobic (FP)~\cite{Hsieh:2010zr, Cao:2012ng,  Chivukula:2006cg, Barger:1980ix, Barger:1980ti}, un-unified (UU)~\cite{Georgi:1989ic, Georgi:1989xz}, and non-universal (NU)~\cite{Li:1981nk, Muller:1996dj, Malkawi:1996fs, He:2002ha, Berger:2011xk}.  For the symmetry breaking pattern $SU(2)_L \times SU(2)_R \times U(1)_{B-L} \to SU(2)_L \times U(1)_Y \to U(1)_{\rm em}$, a new scalar $\Delta$ is needed to realize the first step of symmetry breaking $SU(2)_R \times U(1)_{B-L} \to U(1)_Y$. There are typically two possible choices: the doublet $\Delta\sim(1,2,1/2)$ and triplet $\Delta\sim(1,3,1)$.

In this work we examine the possibility to accommodate the diphoton excess in the $G221$ models with above particular symmetry breaking pattern. We identify the neutral component $\delta^0$ of the triplet or the doublet $\Delta$ as the 750 GeV resonance. A large diphoton decay width of $\delta^0$ is possible due to contribution from $W'$, (doubly) charged scalars in the spectrum. But there is no new colored particle to generate an effective coupling of $S$ to the gluon. This motivates a connection between the excess and the UV completion of $G221$ models. It has been suggested\cite{Bertolini:2013vta, Bertolini:2012im}~\cite{Aydemir:2015oob,Aydemir:2016qqj} that in $SO(10)$ grand unification light colored particles, either scalar or vectorlike fermion, may exist in certain symmetry breaking pattern. With several examples,
we find it is possible to get the demanded diphoton signal with these light states together with the particle content in $G221$ models. There are already some recent papers studying the diphoton excess in left-right model or $SO(10)$ unification~\cite{Dey:2015bur,Dasgupta:2015pbr,Deppisch:2016scs,Berlin:2016hqw,Hati:2016thk,Aydemir:2016qqj}.
But here we explore various $G221$ models. In particular, we focus on the LP and FP models, where a light $W'$ and/or $\delta^{++}$ could play a significant role in the diphoton signature while satisfying various collider constraints. Furthermore, we extend the $G221$ models to the Pati-Salam and $SO(10)$ models, and highlight the connection between the particle content in $G221$ models and possibly light colored particles in $SO(10)$ models.
Given this bottom-up setup, it should be straightforward to work out the diphoton signature in a specific $SO(10)$ model.

This paper is organized as follows. We review the basics of G221 models in Sec.\ref{sec:model}. In Sec.\ref{sec:spec}, we discuss collider searches and constraints on $G221$ models, which are crucial for the prediction of diphoton signal. In Sec.\ref{sec:UV}, we review some possible UV completions of the $G221$ models:  Pati-Salam and $SO(10)$ realizations. For certain symmetry breaking pattern, it is possible to have light colored scalars ( associated with symmetry breaking) or vectorlike fermions in the spectrum. With this, in Sec.\ref{sec:explain}, we discuss the diphoton excess in the LP and FP models with additional light colored scalars or vectorlike fermion. We conclude in Sec.\ref{sec:concl}.


\section{The $SU(2) \times SU(2) \times U(1)$ Model}
\label{sec:model}

In the effective theory framework, the $G221$ models~\cite{Hsieh:2010zr,Cao:2012ng} are the minimal extension of the SM gauge group which incorporates both the $W^\prime$  and $Z^\prime$ bosons.
The $G221$ models have gauge structure $SU(2)\times SU(2)\times U(1)$.
There are two kinds of breaking patterns,
\begin{eqnarray}
\textrm{I}:&&\,\,SU(2)_L \times SU(2)_R \times U(1)_{B-L} \to SU(2)_L \times U(1)_Y \to U(1)_{\rm em}\nonumber\\
\textrm{II}:&&\,\,SU(2)_{L1} \times SU(2)_{L2} \times U(1)_Y \to SU(2)_L \times U(1)_Y \to U(1)_{\rm em}
\end{eqnarray}
%
We will focus on the breaking pattern I in this work.
Depending on fermion assignments, the breaking pattern I includes left-right (LR)~\cite{Mohapatra:1974gc, Mohapatra:1974hk, Mohapatra:1980yp},
lepto-phobic (LP), hadro-phobic (HP), fermio-phobic (FP)~\cite{Hsieh:2010zr, Cao:2012ng,  Chivukula:2006cg, Barger:1980ix, Barger:1980ti} models.

\begin{table}[h!]
\caption{
Possible Higgs multiplets in the breaking pattern I of $G221$ models.}
\begin{center}
\begin{tabular}{|c|c|c|}
\hline
Model & Rep. & Multiplet and VEV
\\
\hline
\begin{minipage}{1.2in}
LR-T, LP-T \\
HP-T, FP-T
\end{minipage} &
$\Delta \sim (1,3,1)$ &
$\Delta=\frac{1}{\sqrt{2}}
\left(\begin{array}{cc}\delta^{+} &\sqrt{2}\delta^{++} \\ \sqrt{2}\delta^{0} & -\delta^{+} \end{array}\right),
\
\VEV{\Delta} = \frac{1}{\sqrt{2}}
\left(\begin{array}{cc} 0 & 0 \\ u & 0 \end{array}\right)$
\\
\hline
\begin{minipage}{1.2in}
LR-T, LP-T \\
HP-T, FP-T
\end{minipage} &
$\Phi \sim (2,\overline{2},0)$ &
$\Phi = \left(\begin{array}{cc}h_1^{0} & h_1^{+} \\ h_2^{-} & h_2^{0} \end{array}\right)$,
$\VEV{\Phi} = \frac{v}{\sqrt{2}}
\left(\begin{array}{cc} c_{\beta} & 0 \\ 0 & s_{\beta} \end{array}\right)$
\\
\hline
\hline
\begin{minipage}{1.2in}
LR-D, LP-D \\
HP-D, FP-D
\end{minipage} &
$\Delta \sim (1,2,\frac{1}{2})$ &
$\Delta =  \left(\begin{array}{c} \delta^+ \\ \delta^0\end{array}\right),\
\VEV{\Delta} =
\frac{1}{\sqrt{2}}
\left(\begin{array}{c}  0 \\ u \end{array}\right)$
\\
\hline
\begin{minipage}{1.2in}
LR-D, LP-D \\
HP-D, FP-D
\end{minipage} &
$\Phi \sim (2,\overline{2},0)$ &
$\Phi=\left(\begin{array}{cc} h_1^{0} & h_1^{+} \\ h_2^{-} & h_2^{0} \end{array}\right)$,
$\VEV{\Phi} = \frac{v}{\sqrt{2}}
\left(\begin{array}{cc} c_{\beta} & 0 \\ 0 & s_{\beta} \end{array}\right)$
\\
\hline
\end{tabular}
\label{tb:Higgs-stage1}
\end{center}
\end{table}

To break the $G221$ symmetry spontaneously, several Higgs multiplets are introduced.
Here in Tab.~\ref{tb:Higgs-stage1} we list possible Higgs multiplets in the breaking pattern I.
At the TeV scale $u$, the breaking $SU(2)_{R} \times U(1)_{B-L} \rightarrow U(1)_{Y}$ is realized by either a scalar doublet  $(1,2,1/2)$ or a scalar triplet $(1,3,1)$~\footnote{The quantum number assignment is under $SU(2)_L\times SU(2)_R\times {U(1)_{B-L}}$}.
At the electroweak scale $v$, the bidoublet scalar is introduced to have subsequent $SU(2)_{L} \times U(1)_{Y} \rightarrow U(1)_{Q}$.
The relevant Lagrangian is
\bea
	{\mathcal L}_{\rm scalar } =
	{\rm Tr} D_\mu \Delta^\dagger D^\mu \Delta + {\rm Tr}  D_\mu \Phi^\dagger D^\mu \Phi
	+ V(\Phi, \Delta),
\eea
where the covariant derivatives are
\bea
      D^\mu\Delta &=&\partial^\mu\Delta
      +\frac{1}{2}ig\left[ \vec{\tau}\cdot\vec{W}_R^\mu \Delta
        -\Delta \vec{\tau} \cdot\vec{W}_R^\mu  \right]
      +\frac{1}{2}ig'B^\mu\, \Delta \nn \\
      D^\mu\Phi&=&\partial_\mu\Phi
      +\frac{1}{2}ig(\vec{\tau}\cdot\vec{W}_L^\mu\Phi
      -\Phi\vec{\tau}\cdot\vec{W}_R^\mu).
\eea
The general scalar potential~\cite{Deshpande:1990ip} is
\bea
	V(\Phi, \Delta)
	&=& -\mu_1^2 \, \mbox{Tr} (\Phi^\dagger \Phi ) \,
- \, \mu_2^2 \left[ \mbox{Tr} (\tilde{\Phi} \Phi^\dagger ) +
 \mbox{Tr}(\tilde{\Phi}^\dagger \Phi ) \right] \, - \, \mu_3^2
 \mbox{Tr}(\Delta \Delta^\dagger )  \nn \cr
 &+ &
 \lambda_1 \, \left[ \mbox{Tr}(\Phi \Phi^\dagger ) \right]^2 \,
 + \,  \lambda_2 \, \left\{ \left[ \mbox{Tr}(\tilde{\Phi}
 \Phi^\dagger ) \right]^2 +
 \left[ \mbox{Tr}(\tilde{\Phi}^\dagger \Phi ) \right]^2 \right\} \, + \,
 \lambda_3 \, \left[ \mbox{Tr}(\tilde{\Phi} \Phi^\dagger )
 \mbox{Tr}(\tilde{\Phi}^\dagger  \Phi)\right]   \nn   \\
  &+  & \lambda_4 \, \left\{ \mbox{Tr}(\Phi^\dagger \Phi )
\left[ \mbox{Tr}(\tilde{\Phi} \Phi^\dagger ) +
 \mbox{Tr}(\tilde{\Phi}^\dagger \Phi ) \right]  \right\} \, + \,
 \rho_1 \,    \left[ \mbox{Tr}(\Delta
 \Delta^\dagger ) \right]^2  \, + \,
\rho_2 \,   \mbox{Tr}(\Delta \Delta )
 \mbox{Tr}(\Delta^\dagger \Delta^\dagger ) \nn\\
 &+ & \alpha_1 \,   \mbox{Tr}(\Phi^\dagger \Phi )
 \mbox{Tr}(\Delta \Delta^\dagger )    \, + \,
\left[ \alpha_2 \,   \mbox{Tr}(\tilde{\Phi}^\dagger \Phi )
\mbox{Tr}(\Delta \Delta^\dagger )  \, + \, h.c. \right]
 \, + \, \alpha_3 \mbox{Tr} ( \Phi^\dagger \Phi
\Delta \Delta^\dagger ) .
\eea
This expression is valid for both doublet and triplet $\Delta$.
The VEVs of the scalar multiplets are defined in Tab.~\ref{tb:Higgs-stage1}.
We also define a quantity $x$, which is the ratio of the VEVs
\be
x \,\, = \,\, \frac{u^2}{v^2} \,\,,
\ee
with $x \, \gg \, 1$.

The gauge couplings for $SU(2)_L$, $SU(2)_R$, and $U(1)_{X}$ are denoted by $g_L$, $g_R$ and $g_{BL}$ with
\bea
g_L = \frac{e}{\sin\theta}, \,\,\,\, g_R = \frac{e}{\cos\theta \sin{\phi}}, \,\,\,\, g_{BL} = \frac{e}{\cos\theta \cos{\phi}} \,\,,
\eea
where the $\theta$ is the SM weak mixing angle and the $\phi$ is denoted the new mixing angle between $SU(2)_R$ and $U(1)_{B-L}$.
We denote the gauge bosons in $G221$ models as
\begin{align}
SU(2)_L&: W_{1,\mu}^{\pm}, W_{1,\mu}^{3},\nonumber\\
SU(2)_R&: W_{2,\mu}^{\pm}, W_{2,\mu}^{3},\nonumber\\
U(1)_{B-L}&: X_{\mu}.
\end{align}
Both $W^\prime$ and $Z^\prime$ bosons obtain masses
and mix with the SM gauge bosons after symmetry breaking.
To order $1/x$, the eigenstates of the charged gauge bosons are
\begin{eqnarray}
W_\mu^\pm &=& {W_1^\pm}_\mu +\frac{\sin\phi \sin2\beta}{x
\tan\theta}{W_2^\pm}_\mu \, ,\\
{W^\prime}_\mu^\pm &=&  -\frac{\sin\phi \sin2\beta}{x \tan\theta}
{W_1^{\pm}}_{\mu}+{W_2^{\pm}}_{\mu}  \, .
\end{eqnarray}
While the eigenstates of the neutral gauge bosons are
\begin{eqnarray}
Z_\mu &=& {W_Z^3}_\mu +\frac{\sin \phi \cos^3 \phi}{x\sin \theta}
                           {W_H^3}_\mu\, ,\\
Z_\mu^{\prime} &=&  -\frac{\sin\phi\cos^3\phi}{x\sin\theta}
{W_Z^3}_\mu + {W_H^3}_\mu \, ,
\end{eqnarray}
where $W_H^3$ and $W_Z^3$ are defined as
\begin{eqnarray}
{W_H^3}_{\mu} &=& \cos\phi   {W_2^3}_{\mu} - \sin\phi X_{\mu}\,,\\
{W_Z^3}_{\mu} &=& \cos\theta {W_1^3}_{\mu} -\sin\theta (\sin\phi {W_2^3}_{\mu} + \cos\phi X_{\mu})\,,\\
{A    }_{\mu} &=& \sin\theta {W_1^3}_{\mu} +\cos\theta (\sin\phi {W^3_2}_{\mu} + \cos\phi X_{\mu}).
\end{eqnarray}
We also obtain
the masses of the $W^{\prime}$ and $Z^{\prime}$.
For the doublet scalar, the masses are
\bea
m_{W^{\prime}}^{2} = \frac{e^{2}v^{2}}{4\cos^{2}\theta \sin^{2}{\phi}}\left(x+1\right)\,,
\quad
m_{Z^{\prime}}^{2}  = \frac{e^{2}v^{2}}{4\cos^{2}\theta\sin^{2}{\phi}\cos^{2}{\phi}}\left(x+\cos^{4}{\phi}\right)\,,
\label{mzp_bp1}
\eea
and for the triplet scalar we have
\bea
m_{W^{\prime}}^{2} = \frac{e^{2}v^{2}}{4\cos^{2}\theta \sin^{2}{\phi}}\left(2x+1\right)\,,
\quad
m_{Z^{\prime}}^{2}  = \frac{e^{2}v^{2}}{4\cos^{2}\theta\sin^{2}{\phi}\cos^{2}{\phi}}\left(4x+\cos^{4}{\phi}\right)\,.
\label{mzp_bp1t}
\eea

We would like to identify the neutral component $\delta^0$ of the triplet $\Delta$ or the doublet $\Delta$ as the 750 GeV resonance,
\bea\label{eq:Sdelta0}
S \equiv \sqrt{2}{\rm Re}\delta^0\,.
\eea
The $S$ should have very small tree-level decay branching ratios to the $WW$ and $ZZ$ final states.
This gives rise to small mixing between $W$ and $W'$, and small mixing between $Z$ and $Z'$.
This could be realized by taking  the following limits 
\bea\label{eq:smixing}
	\cos\phi \to 0, \quad \sin2\beta \to 0\,.
\eea
According to (\ref{mzp_bp1}), (\ref{mzp_bp1t}), this implies a mass hierarchy $m_{W'} \ll m_{Z'}$. As we will discuss later, this hierarchy is crucial to accommodate diphoton excess in LP and FP models that satisfies the collider constraints on $W', Z'$.

In the scalar sector, there are neutral scalars $( h_1, h_2, \delta^0)$, charged scalars $(h_1^\pm, h_2^\pm, \delta^\pm )$, and double charged scalars $(\delta^{++}, \delta^{--})$ in triplet case.
In the limit of small mixings between $\delta^0$ and $(h_1, h_2)$, the mass of the neutral component of the $\Delta$ is
\bea
	m_{S}^2 = 2 \rho_1  u^2,
\eea
for the triplet scalar multiplet, and
\bea
	m_{S}^2 = 2 (\rho_1 + \rho_2)  u^2,
\eea
for the doublet case ($\rho_1, \rho_2$ define the same operator).
The doubly charged  scalar mass is
\bea
	m_{\delta^{++}}^2 = 2 \rho_2 u^2.
\eea
For the charged scalars, only one combination of $h_1^\pm, h_2^\pm, \delta^{\pm}$ is left in the physical spectrum after
symmetry breaking. In small mixing limit, its coupling to $S$ is negligible.

In the triplet case, the $S W'W'$ and $S Z'Z'$ couplings are
\begin{eqnarray}
 \lambda_{SW'}=g^{\mu\nu}\frac{e^2 u}{2\sin^2\theta}  f_{S W'W'}, \quad
 \lambda_{SZ'}=g^{\mu\nu}\frac{e^2 u}{2\sin^2\theta\cos^2\theta}  f_{S Z'Z'},
\end{eqnarray}
with the dominant coupling strengths are
\begin{eqnarray}
  f_{S W'W'} = \frac{\tan^2\theta}{\sin^2\phi},\quad
  f_{S Z'Z'} = \frac{\sin^2\theta }{\sin^2\phi\cos^2\phi}.
\end{eqnarray}
The relevant triple and quartic gauge boson couplings are
\bea
	 W'W' Z: && \frac{e \sin\theta}{\cos\theta}  \left( 1- \frac{\cos^4\phi}{2x \sin^2\theta}\right), \\
	 W'W'Z \gamma: && \frac{e^2 \sin\theta}{\cos\theta}  \left( 1- \frac{\cos^4\phi}{2x \sin^2\theta}\right),
\eea
with the same lorentz structures as $WWZ$ and $WWZ\gamma$.
In the doublet case, we have similar Feynman rules with $x \to x/2$.
The relevant triple scalar coupling is
\bea
	\lambda_{S\delta^{++}}= 2 (\rho_1 + 2 \rho_2) u.
\eea
The above Feynman rules will be used for calculating the diphoton rate.
%


\section{Collider Searches on $G221$ Models}
\label{sec:spec}

The $W'$ and $Z'$ gauge bosons as predicted by $G221$ models are subject to the constraints
from the LHC searches.
If the $Z'$ and $W'$ decay leptonically,
the LHC searches place the most stringent constraints on their masses.
The current searches for the high mass resonances on the lepton plus transverse missing energy final states set the limit on the  $W'$ mass: $m_{W'} > 2$ TeV~\cite{ATLAS:2014wra}.
Similarly, the dileptonic searches put tightest constraints on $Z'$ mass: $m_{Z'} > 2.9$ TeV~\cite{Aad:2014cka}.

Among various possibilities of $G221$ models, $W'$ couples to the leptons in the LR, HP models.
So $W'$ is expected to be around 2 TeV in these cases, which could not significantly contribute to the diphoton signal.
This pushes us to think about the LP and FP models, which forbid leptonic decay of $W'$.
The fermion contents of these two models are shown in Tab.~\ref{tb:models}.
In the FP model, all SM quarks and leptons are $SU(2)_R$ singlets and the $W_R$ cannot decay to any SM fermions.
In the LP model, the right-handed neutrinos could be several TeV for the seesaw neutrino masses.
We could realize the LP model without gauge anomaly by integrating out the heavy right-handed neutrinos.
%
%
%
%
%
%

\begin{table}[h]
\begin{center}
\caption{
The charge assignments of the SM fermions under
the leptophobic $G221$ models.
}
\label{tb:models}
\vspace{0.125in}
\begin{tabular}{|c|c|c|c|}
\hline Model & $SU(2)_L$ & $SU(2)_R$ & $U(1)_{B-L}$ \\
\hline
\hline
Lepto-phobic &
$\left(\begin{array}{c} u_L \\ d_L \end{array}\right),\left(\begin{array}{c} \nu_L \\ e_L \end{array}\right)$ &
$\left(\begin{array}{c} u_R \\ d_R \end{array}\right),\left(\begin{array}{c} {\mathcal N}_R \\ e_R \end{array}\right)$ &
$ {\textrm{all fermions}} $
\\
\hline
Fermio-phobic &
$\left(\begin{array}{c} u_L \\ d_L \end{array}\right),\left(\begin{array}{c} \nu_L \\ e_L \end{array}\right)$ &
 &
$ {\textrm{SM fermions}} $\\
\hline
\end{tabular}
\end{center}
\end{table}

Therefore, the $W'$ in the LR, HP models can evade the stringent bound from direct search in the lepton plus missing energy final states.
%
%
However, the $W'$ cannot be too light, due to the electroweak precision test~\cite{Hsieh:2010zr,Cao:2012ng} and anomalous gauge coupling. These set a lower bound $m_{W'} > 300$ GeV.
In the LP model, the $W'$ can be searched by di-jet final states as well.
Both Tevatron and LHC set limits on the di-jet resonances.
For the $W'$ lighter than 1 TeV, the constraints from Tevatron are stronger.
As shown in Ref.~\cite{Aaltonen:2008dn}, given the SM like coupling of the $W'$,
the cross section is only a little above the exclusion limits.
Therefore, depending on the $W'$ coupling to the SM quarks, the $W'$ boson could be allowed at several hundred GeV region.
It is also possible that the $W'$ decays to $WZ$ final states, which have been searched at the LHC~\cite{Aad:2015ipg}.
Given  the assumptions of a small $W'WZ$ coupling in the LP model~\footnote{Of course, if the $W'WZ$ coupling is not so small due to possibly large $\cos\phi$, it could induce the diboson signatures in the LP model~\cite{Gao:2015irw}. },
the branching ratio $W' \to WZ$ would be highly suppressed and the current limits on the $W' \to WZ$ are too weak to be relevant.
Regarding to the flavor constraints, the light $W'$
could contribute to the box diagrams in the neutral meson mixing system, such as
$K-\bar{K}$ mixing, $B-\bar{B}$ mixing.
Given the general right-handed flavor mixing, the constraints from $K-\bar{K}$ mixing  are
quite tight and $W'$ is required to be several TeV~\cite{Blanke:2011ry}.
However, if the right-handed CKM matrix takes special form, these constraints could be evaded~\cite{Buras:2010pz} and a light $W'$ with several hundred GeV mass is still allowed.
%
%
In the FP model, the $W'$ does not couple to the SM fermions, and so escapes the di-jet and flavor constraints.
%
%
Therefore, a light $W'$ is still possible, and we may expect it contributes significantly to the diphoton signature.

There is another caveat in LP and FP models.
Due to the $Z-Z'$ mixing, the $Z'$ could still decay to dilepton, and thus the $Z'$ needs to be heavy with around 2 TeV mass.
This implies a mass hierarchy for new gauge bosons: several hundred GeV $W'$ and several TeV $Z'$.
According to the mass relations, this could be realized in the parameter region $\cos\phi \to 0$,
which is consistent with the requirement of small $W, W'$ and $Z, Z'$ mixing in (\ref{eq:smixing}) for the diphoton excess.

In the $G221$ models, the right-handed triplet or doublet $\Delta$ could be around TeV scale.
With negligible small mixing between $\Delta$ and the bidoublet $\Phi$, the only charged scalar that couples to the $\Delta$ neutral component is $\delta^{++}$ in the triplet case. $\delta^{++}$ can be pair produced on hadron collider via s-channel photon exchange. Depending on the decay modes, the LHC searches put constraints on lower mass bound of $\delta^{++}$. We find $m_{\delta^{++}} > 374\,(438)\,$GeV if it mainly decays to $e^{\pm}e^{\pm}$ ($\mu^{\pm}\mu^{\pm}$)~\cite{ATLAS:2014kca}. Due to the small mixing of $W, W'$, the vector-boson fusion production of $\delta^{++}$ or its decay into $W^\pm W^\pm$ is negligible.
Therefore, there are still enough parameter space that the $\delta^{++}$ could be light and may play a role in diphoton excess.


\section{Possible UV Completion from SO(10) Grand Unification}
\label{sec:UV}

%
The  $G221$ models could be incorporated into the Pati-Salam $SU(4)_c \times SU(2) \times SU(2)$~\cite{Pati:1974yy, Pati:1973uk} or $SO(10)$ groups~\cite{Georgi:1974sy, Fritzsch:1974nn}.
The $SO(10)$ grand unified theory (GUT) provides us a very attractive framework.
Among different breaking patterns for a $SO(10)$ group,
there are many cases that $SU(2)_L \times SU(2)_R \times U(1)_{B-L}$ is an intermediate step, which then break down to the SM gauge group via the breaking pattern I.

As we discussed in $G221$ models, the breaking pattern I can be realized by either a right-handed triplet or doublet. In context of $SO(10)$ symmetry, the triplet $\Delta$ can be embedded in $\textbf{126}$ representation as the following,
\bea
\Delta\equiv (1, 1, 3)_2 \subset (\overline{10}, 1, 3) \subset 126\,,
\eea
where $(1, 1, 3)_2$ are the charges under $SU(3)_c\times SU(2)_L\times SU(2)_R\times U(1)_{B-L}$ and $(\overline{10}, 1, 3)$ are the charges under $SU(4) \times SU(2) \times SU(2)$.
The advantage of the $\textbf{126}$ representation is that it provides a Majorana mass term for the right-handed neutrino.
The right-handed doublet can be embedded in $\textbf{16}$ representation as the following,
\bea
\Delta\equiv (1, 1, 2)_1 \subset (\overline{4}, 1, 2) \subset 16\,.
\eea
In $G221$ models, a bidoublet $\Phi$ is chosen for the subsequent breaking of the SM gauge symmetry. It can be embedded in $\textbf{10}$ representation of $SO(10)$ symmetry. So the breaking pattern can be summarized as follows,
\bea
	SU(3)_c  \times G_{221} \xrightarrow{\left<\mathbf{ 126} \, {\textrm{or}}\,  \mathbf{ 16}\right>} {\rm SM}
	\xrightarrow{\left<\mathbf{ 10}\right>} SU(3)_c \times U(1)_{\rm Q}.
\eea

There are many possibilities to break the original $SO(10)$ symmetry~\cite{Hartmann:2014fya}.
Among those, we list the possible symmetry breaking patterns involving the $G_{221}$ symmetry,
\bea
	&& SO(10) \xrightarrow{\left<\mathbf{ 210}\right>} G_{422}
	\xrightarrow{\left<\mathbf{ 45}\right>} SU(3)_c  \times G_{221}, \\
	&& SO(10) \xrightarrow{\left<\mathbf{ 54}\right>} G_{422} \times P
	\xrightarrow{\left<\mathbf{ 210}\right>} SU(3)_c  \times G_{221}  \times P, \\
	&& SO(10) \xrightarrow{\left<\mathbf{ 54}\right>} G_{422} \times P
	\xrightarrow{\left<\mathbf{ 45}\right>} SU(3)_c  \times G_{221} , \\
	&& SO(10) \xrightarrow{\left<\mathbf{ 210}\right>} SU(3)_c  \times G_{221}  \times P
	\xrightarrow{\left<\mathbf{ 45}\right>} SU(3)_c  \times G_{221}  ,
\eea
and
\bea
	&& SO(10) \xrightarrow{\left<\mathbf{ 45}\right>}  SU(3)_c  \times G_{221}  \times P
	\xrightarrow{\left<\mathbf{ 45}\right>}  SU(3)_c  \times G_{211} , \\
	&& SO(10) \xrightarrow{\left<\mathbf{ 210}\right>}  SU(3)_c  \times G_{221}  \times P
	\xrightarrow{\left<\mathbf{ 45}\right>} SU(3)_c  \times G_{211}.
\eea
Here $ G_{422}$ is the Pati-Salam symmetry $SU(4) \times SU(2) \times SU(2)$, $G_{211}$ denotes
$SU(2)_L \times U(1)_R \times U(1)_{B-L}$, and $P$ is the parity between the left-right symmetry.
For some symmetry breaking patterns,
it is possible to have light colored scalars or colored vectorlike fermions. We will discuss these possibilities in detail.

\subsection{Light Colored Scalars}

To break the $SO(10)$ symmetries, several scalar multiplets are introduced, as shown in above breaking patterns.
The possibilities that certain components of scalar multiplets are much lighter than others, i.e. below the TeV scale, have been discussed in literatures.
%
%

In Ref.~\cite{Bertolini:2013vta, Bertolini:2012im}, the possible symmetry breaking patterns are
\bea
	&& SO(10) \xrightarrow{\left<\mathbf{ 45}\right>}  SU(3)_c  \times G_{221}
	\xrightarrow{\left<\mathbf{ 126}\right>} {\rm SM}
	\xrightarrow{\left<\mathbf{ 10}\right>} SU(3)_c \times U(1)_{\rm Q}, \\
	&& SO(10) \xrightarrow{\left<\mathbf{ 45}\right>}  SU(3)_c  \times G_{221}
	\xrightarrow{\left<\mathbf{ 45}\right>} SU(3)_c  \times G_{211}
	\xrightarrow{\left<\mathbf{ 126}\right>} {\rm SM}
	\xrightarrow{\left<\mathbf{ 10}\right>} SU(3)_c \times U(1)_{\rm Q}.
\eea
They considered a light color octet from the $\textbf{126}$ representation as follows
%
\bea
(8, 2)_{1/2} \subset (8, 2, 2)_0 \subset 126.
\eea
where $(8, 2)_{1/2}$ is the quantum charge under the SM gauge group and $(8, 2, 2)_0$ is the charge under $SU(3)_c\times G_{221}$.
The color octet can couple to SM diquarks via interaction $16\,\overline{126}\,16$. This yukawa coupling is not directly related to the SM yukawa.

In Ref.~\cite{Aydemir:2016qqj}, another symmetry breaking pattern has been considered
\bea
	SO(10) \xrightarrow{\left<\mathbf{ 210}\right>} G_{422}
	\xrightarrow{\left<\mathbf{ 210}\right>} SU(3)_c  \times G_{221}
	\xrightarrow{\left<\mathbf{ 126}\right>} {\rm SM}
	\xrightarrow{\left<\mathbf{ 10}\right>} SU(3)_c \times U(1)_{\rm Q}.
\eea
The light scalar is now a color triplet and also belongs to the $\textbf{126}$ representation
\bea
	(\bar{3}, 1)_{4/3} \subset (\bar{3}, 1, 3)_{2/3} \subset (\overline{10}, 1, 3) \subset 126,
\eea
where $(\overline{10}, 1, 3)$ are the quantum charges under $G_{422}$, $(\bar{3}, 1, 3)_{2/3}$ are the charges under $SU(3)_c\times G_{221}$, and $(\bar{3}, 1)_{4/3}$ are the charges under the SM gauge group.
In their setup, the colored scalar is taken to be as light as several hundred GeV.
It is likely that all three scalars in $(\bar{3}, 1, 3)_{2/3}$ that fall apart in $G221$ breaking down to the SM have the same scale.
Furthermore, in PS symmetry breaking, there is another color sextet $(\bar{6},1,3)_{-2/3}$ deduced from $(\overline{10}, 1, 3)$.
The possibility that both color triplet and sextet are light is studied in $SO(10)$ GUT as well~\cite{Aydemir:2015oob}.
Via yukawa interaction $16\,\overline{126}\,16$, the triplet scalars could couple to a SM lepton and a quark, and the sextet couples to diquark.

Let us denote the light colored scalar as $\chi$. The general couplings between
the scalar $\Delta$ and $\chi$ is written as
\bea\label{eq:rho3}
	{\mathcal L}  \supset  \rho_3\chi^\dagger \chi \textrm{Tr}(\Delta\Delta^\dagger),
\eea
for both doublet and triplet.
We will utilize this Lagrangian to calculate the diphoton rate.

\subsection{Light Vectorlike Fermions}

The minimal $SO(10)$ GUT model could be extended to
incorporate vectorlike fermions  as $16 \oplus \overline{16}$~\cite{Babu:1994pd, Babu:2002fs} and $10 \oplus \overline{10}$ of the $SO(10)$ representation.
The $SO(10)$ representations $16$ and $10$ can be decomposed to $G_{422} \to SU(3)_c \times G_{221}$ as
\bea\label{eq:dec}
16 &=& (4, 2, 1) + (\bar{4}, 1, 2) = (3,2,1)_{+1/3}+(1,2,1)_{-1} + (\bar{3},1,2)_{-1/3} + (1,1,2)_{+1},\nn\\
10 &=& (6,1,1) + (1, 2,2) = (3,1,1)_{-2/3} + (\bar{3},1,1)_{+2/3} + (1,2,2)_0.	
\eea
The vectorlike fermions in $16 \oplus \overline{16}$ and $10 \oplus \overline{10}$
\bea
 16_F = \left( {\mathcal Q} \,|\, \bar{\mathcal Q'} \right), \quad {10}_F = \left( {\mathcal D} \,|\,  {\mathcal N} \right),
\eea
where ${\mathcal Q}$ and ${\mathcal Q}'$ transform as $(4,2,1)$ and  $(4,1,2)$ of the symmetry group $G_{422}$, and ${\mathcal D}$ and ${\mathcal N}$ transform as $(6,1,1)$ and $(1, 2,2)$ of the $G_{422}$.
Here ${\mathcal Q}, {\mathcal Q}'$ denote two
complete vectorlike families of quarks and leptons, and
${\mathcal D}, {\mathcal N}$ denote two vectorlike families of down-type quarks and leptons. According to the decomposition in (\ref{eq:dec}), we denote the multiplets in G221 representations as the following,
\bea
{\mathcal Q}_{L,R} = (Q, L)_{L,R}, \quad {\mathcal Q}'_{L,R} = (Q', L')_{L,R}, \quad
{\mathcal D}_{L,R} = (D, \bar{D}')_{L,R}, \quad {\mathcal N}_{L,R} = N_{L,R}\,.
\eea
Given the vectorlike fermion families,
we have following $SO(10)$ Yukawa interactions
\bea
	{\mathcal L}_{\rm Yuk} &\supset&  Y_{10}\, 16\, \overline{10}\, 16 + Y_{120}\, 16\, \overline{120}\, 16 + Y_{126}\, 16\, \overline{126}\, 16
\eea
where we embed the light doublet or triplet scalar $\Delta$
in the $16$ or $126$ representation that breaks the $G_{221}$ symmetry down to the SM gauge symmetry.
If we use the right-handed triplet scalar in the $126$ representation, the Yukawa term is $16_F\,\overline{126}_H\,16_F$.
After symmetry breaking, this Yukawa term only generates coupling of $S$ to the neutral leptons.
No vectorlike fermions contribute to the diphoton signature.
So we choose the right-handed doublet $\Delta$ in the $16$ representation.
The Yukawa term involves in
$16_F$ and $10_F$
\bea
	{\mathcal L}_{\rm Yuk} &\supset&  Y_{10}\, \overline{10}_F\, 16_H\, 16_F + h.c.
\eea
After Pati-Salam symmetry breaking, the Yukawa term for the right-handed doublet reduces to
\bea\label{eq:VLFdelta}
	{\mathcal L}_{\rm Yuk} \supset  y_\Delta \bar{Q}'_L {\Delta} D'_R + y_\Delta {\Delta}^\dag \bar{ N}_L  L_R + h.c.
\eea
%
%
If the vectorlike fermions are around TeV scale, they are expected to contribute to
the diphoton signature.

In summary, depending on the symmetry breaking patterns, light colored scalars or light vectorlike fermions could exist in $SO(10)$ GUT models.
For the right-handed triplet in 126 representation and doublet in 16 representation, the neutral component $S$ can couple to
light colored/charge scalars or vectorlike fermions, which then contribute to the diphoton signature.
%
Here we only pick up several models discussed in literatures. There could be
more $SO(10)$ GUT models which may have light particles in the spectrum.
%
It might also be possible to extend the $SO(10)$ symmetry to $E_6$ group.
Our framework could also extend to SUSY $SO(10)$ GUT framework to solve hierarchy problem.
%


\section{Explanations on Diphoton Excess}
\label{sec:explain}

In this section we examine the diphoton excess in the FP and LP models and their UV completions,
where a light $W'$ with several hundred GeV mass could exist.
We identify the neutral component of the right-handed triplet and doublet $\Delta$ as the 750\,GeV resonance $S$ as in (\ref{eq:Sdelta0}).
The diphoton signature comes from the loop-induced process $gg \to S \to \gamma \gamma$.

\begin{figure*}[t]
\includegraphics[width=7.3cm]{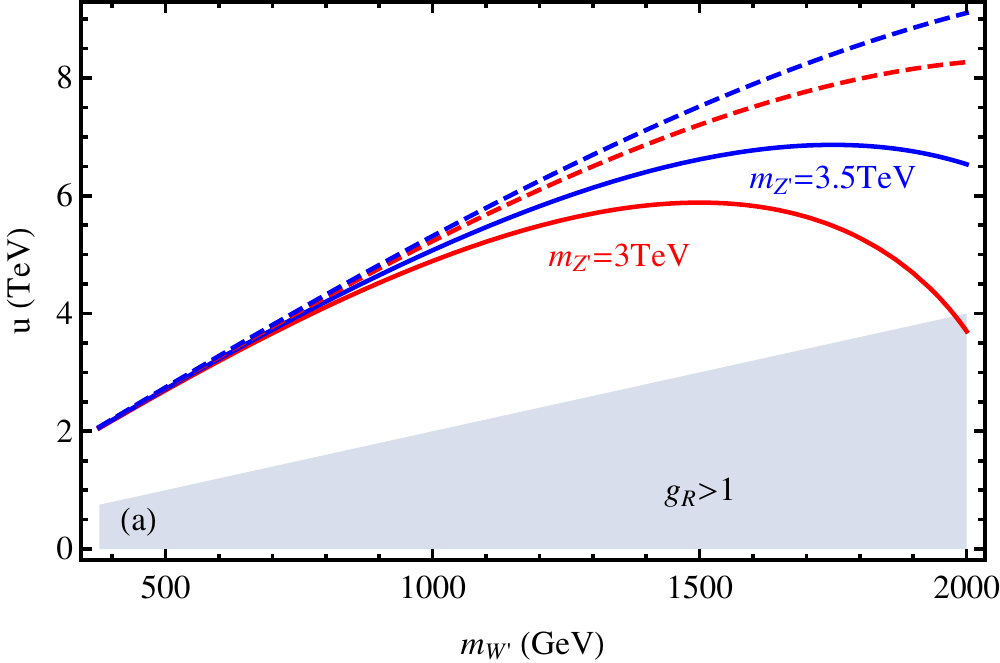}\quad\quad
\includegraphics[width=7.4cm]{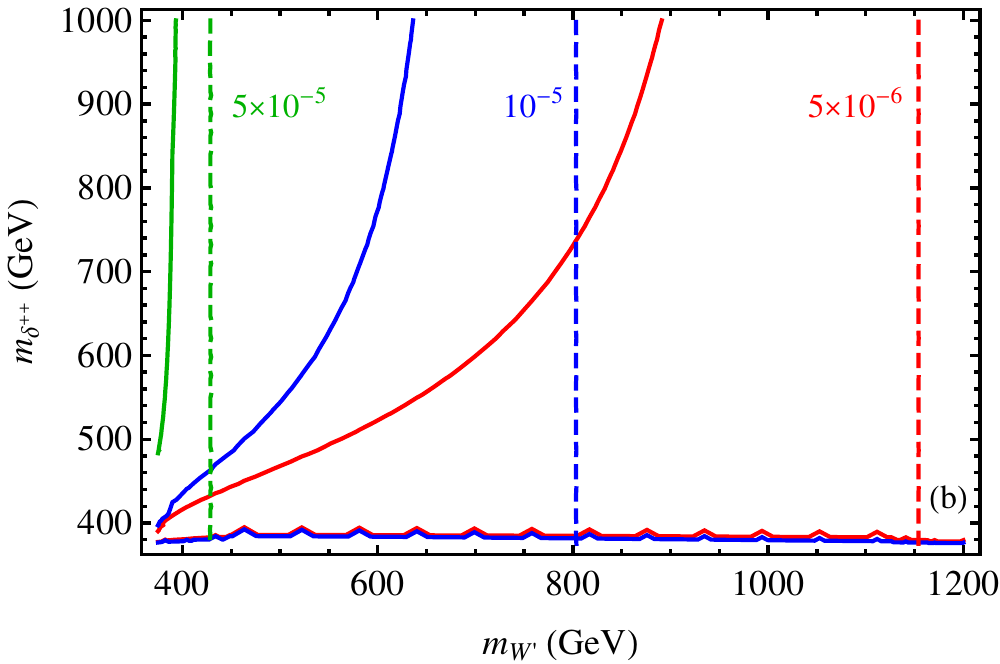}
\caption{(a) $u$ as function of $m_{W'}$ with $m_{Z'}=$\,TeV (red) and  $m_{Z'}=3.5\,$TeV (blue) in doublet case (dash) and the triplet case (solid). Shaded region denotes $g_R>1$. (b) contours of $\Gamma_{\gamma\gamma}$ (GeV) in the doublet case (dash) and the triplet case (solid) on $m_{W'}-m_{\delta^{++}}$ plane assuming $m_{Z'}=3\,$TeV. For the former, there is no $\delta^{++}$ and $\Gamma_{\gamma\gamma}$ only depends on $m_{W'}$.}
\label{fig:umWp}
\end{figure*}

The advantage of the $G221$ models is that there are several charged particles that couple to the $S$ and thus contribute significantly to the diphoton width.
In the doublet case, the $S$ couples to the $W'^\pm$ and a linear combination of $h_1^\pm, h_2^\pm$; in the triplet case, it couples to the doubly charger scalar $\delta^{++}$ in addition. Given the small mixing between $\Delta$ and bidoublet $\Phi$, the contribution from the single charged scalar is always negligible. So the diphoton width is mainly contributed by $W'^\pm$ and $\delta^{++}$ loops, and depends on their masses and triple couplings to $S$.
In $x\gg1$ limit the only relevant triple coupling in doublet case is,
\begin{eqnarray}\label{eq:tripleSD}
\lambda_{SW'}= \frac{2 m_{W'}^2}{u} \,,
\end{eqnarray}
where $\cos\phi\approx m_{W'}/m_{Z'}$, which then determines $g_R$ and $u$. For the triplet case there are two relevant couplings,
\begin{eqnarray}\label{eq:tripleST}
\lambda_{SW'}= \frac{2 m_{W'}^2}{u} \, ,\quad
\lambda_{S\delta^{++}}=
\frac{2 m_{\delta^{++}}^2+m_S^2}{u}\,,
\end{eqnarray}
where $\cos\phi\approx \sqrt{2}m_{W'}/m_{Z'}$. Fig.\ref{fig:umWp} (a) shows $u$ as function of $m_{W'}$ with different $m_{Z'}$ that satisfies the lower bound $m_{Z'}>2.9\,$TeV in both cases.

With (\ref{eq:tripleSD})(\ref{eq:tripleST}), we derive the diphoton width for the doublet cases,
\begin{eqnarray}
\Gamma_{S \rightarrow \gamma \gamma}
 =
  \frac {\alpha^2}{256 \pi^3} \frac{m_S^3}{u^2}
\left|
  A_1(\tau_{W'})
\right|^2,
\end{eqnarray}
and for the triplet case
\begin{eqnarray}
\Gamma_{S \rightarrow \gamma \gamma}
 =
  \frac {\alpha^2}{256 \pi^3} \frac{m_S^3}{u^2}
\left|
  A_1(\tau_{W'}) + 4\left(1+2\tau^{-1}_{\delta^{++}}\right) A_0(\tau_{\delta^{++}})
\right|^2,
\end{eqnarray}
where $\tau_i=4m_i^2/m_S^2$ and loop functions $A_i$ are defined in Appendix \ref{app:decayw}. When $m_{W'},m_{\delta^{++}}>m_S/2$, $A_1, A_0$ are both real but in opposite sign.
Fig.\ref{fig:umWp} (b) presents the contours of $\Gamma_{\gamma\gamma}$ (in unit of GeV) on $m_{W'}-m_{\delta^{++}}$ plane for the doublet case (dash) and the triplet case (solid) respectively, assuming $m_{Z'}=3\,$TeV. As $\Gamma_{\gamma\gamma}$ is suppressed by $1/u$ in both cases, a large diphoton width prefers a light $W'$.
Comparing the two models, we see the destructive interference of $W'$ and $\delta^{++}$ contributions in the triplet case. For generic $m_{\delta^{++}}>m_S/2$, $W'$ contribution is dominant and a lighter $m_{W'}$ is required in the triplet case for the same decay width. The doubly charged scalar $\delta^{++}$ could be important only if $m_{\delta^{++}}$ is very close to $m_S/2$.

\begin{figure*}[t]
\includegraphics[width=7.3cm]{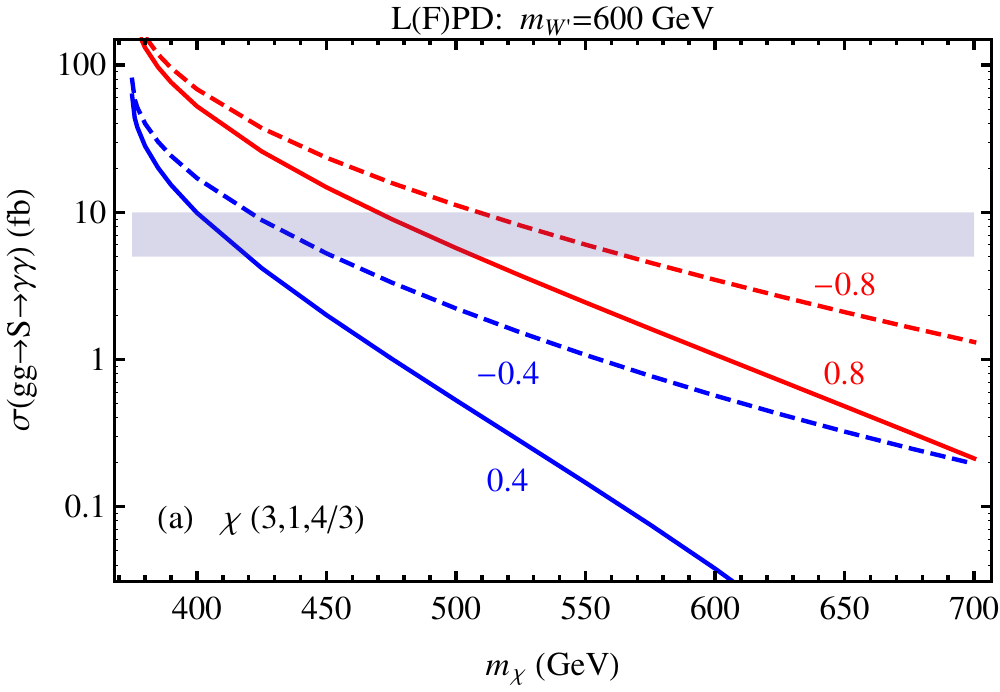}\quad\quad
\includegraphics[width=7.3cm]{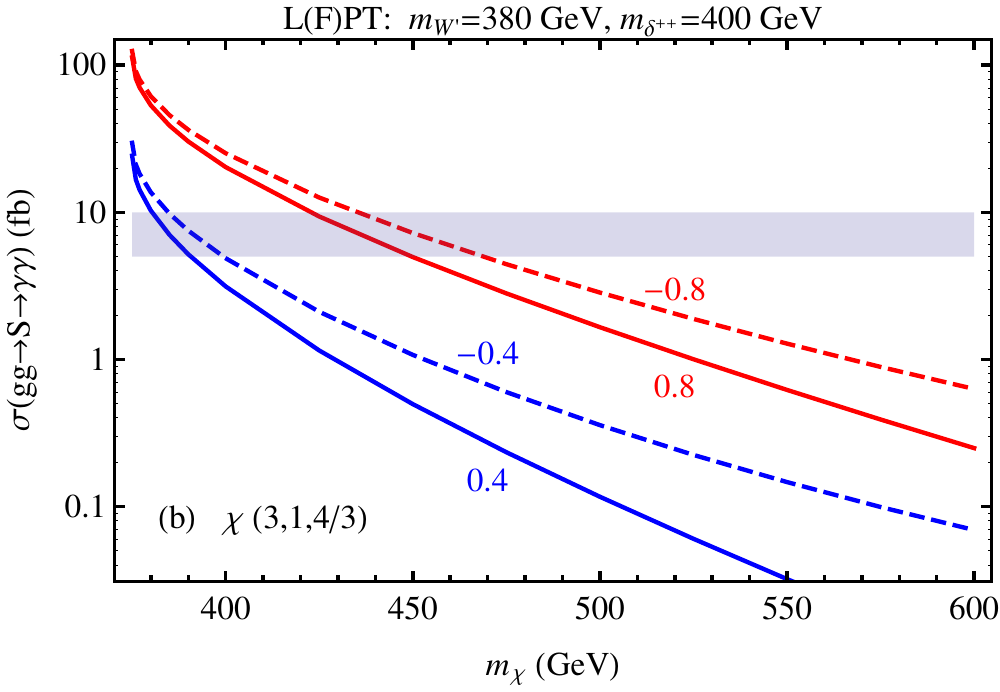}\\
\includegraphics[width=7.3cm]{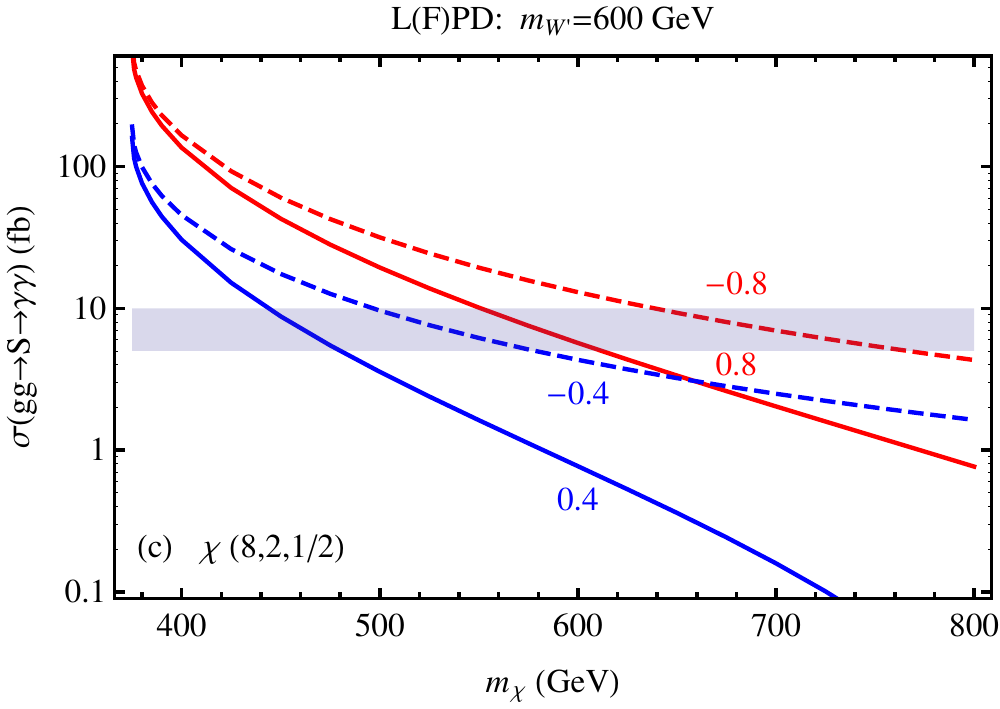}\quad\quad
\includegraphics[width=7.3cm]{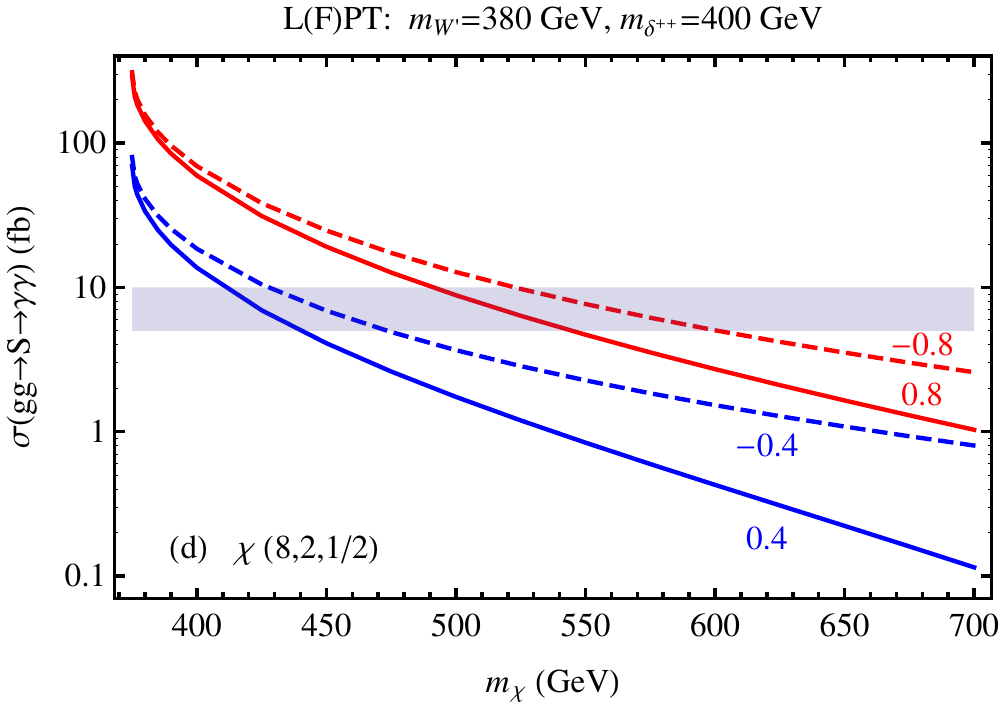}\\
\caption{$\sigma(gg\to S\to\gamma\gamma)$ as function of colored scalar mass $m_\chi$ for (a) L(F)PD with color triplet $\chi$, (b) L(F)PT with color triplet $\chi$, (c) L(F)PD with color octet $\chi$, (d) L(F)PT with color triplet $\chi$. For L(F)PD, we choose different $m_{W'}=600\,$GeV, while for L(F)PT we choose $m_{W'}=380\,$GeV and $m_{\delta^{++}}=400\,$GeV. In each panel, the color lines denote different $\rho_3$.}
\label{fig:Xsecaas}
\end{figure*}

To initiate gluon fusion production of $S$ without large decay branching ratio to SM fermions, new colored particles that couple to $S$ are required. The $G221$ models do not provide such candidates, but as we discussed in Sec.\ref{sec:UV}, light colored particles are expected for certain symmetry breaking pattern of the UV completion of G221 models.

The first possibility is to have new colored scalars below TeV scale. According to (\ref{eq:rho3}), the triple coupling of $S$ and the colored scalar $\chi$ is $\lambda_{S\chi}=\rho_3u$. For illustration, we consider two examples of a light colored scalar in context of $SO(10)$ GUT as we discussed in Sec.\ref{sec:UV}. Under SM gauge group $SU(3)_C\times SU(2)_L\times U(1)_Y$, they are in representations $(\bar{3},1)_{4/3}$ and $(8,2)_{1/2}$ respectively. Hereafter we use L(F)PD for the LP (FP) models with doublet scalar $\Delta$, and L(F)PT for the LP (FP) models with triplet scalar $\Delta$. Fig.\ref{fig:Xsecaas} presents $\sigma(gg\to S\to\gamma\gamma)$ as function of $m_\chi$ for these two samples of colored scalars, assuming $m_{Z'}=3\,$TeV and $\tan\beta=0.01$. In each panel, the grey band shows the range of observed signal. The color lines denote different $\rho_3$, and the solid, dash lines for positive, negative values respectively. Here we choose $m_{W'}$ ($m_{\delta^{++}}$) such that there is considerable interference of $\chi$ with $W'$ ($\delta^{++}$) in diphoton loop. The interference is constructive when $\rho_3<0$. Compared with L(F)PD, a lighter $W'$ is more preferred in L(F)PT due to the negative contribution of $\delta^{++}$. For colored scalars in two different representations, the color octet $(8,2)_{1/2}$ admits a much larger diphoton rate than the color triplet $(3,1)_{4/3}$, and $m_\chi$ could be heavier.

In general a rather light colored scalar is needed, i.e. $m_\chi\sim500\,$GeV. This might be a problem for the color triplet because it may couple to SM quark and lepton and behave like a scalar leptoquark. For a sizable Yukawa that the scalar decays promptly inside the detector, the current leptoquark searches apply. The constraints on $m_\chi$ depend on the decay modes. The triplet $(3,1)_{4/3}$ couples to a charged lepton and a down-type quark, and the lower bound is $m_\chi>740\,(1000)\,$GeV if it decays to $\tau b$ ($e q$) with 100\% branching ratio~\cite{Khachatryan:2014ura, CMS:2014qpa}. This may exclude the interpretation of diphoton excess by $(3,1)_{4/3}$ alone. But it may evade the constraint for certain range of small Yukawa if it decays as a displaced vertex. The probe in this region is still weak. For the color octet, it may couple to two SM quarks, which implies dijet events. There are searches of dijet resonances of several hundred GeV on Tevatron~\cite{Aaltonen:2008dn}. A single color octet can be produced by quark-initiated production. But the cross section is highly model-dependent and might be suppressed by a small Yukawa. The pair production rate is fixed by the gluon coupling, but the four-jet final states suffer from the overwhelming QCD backgrounds. So the constraint on color octet could be fairly weak.

\begin{figure*}[t]
\includegraphics[width=7.3cm]{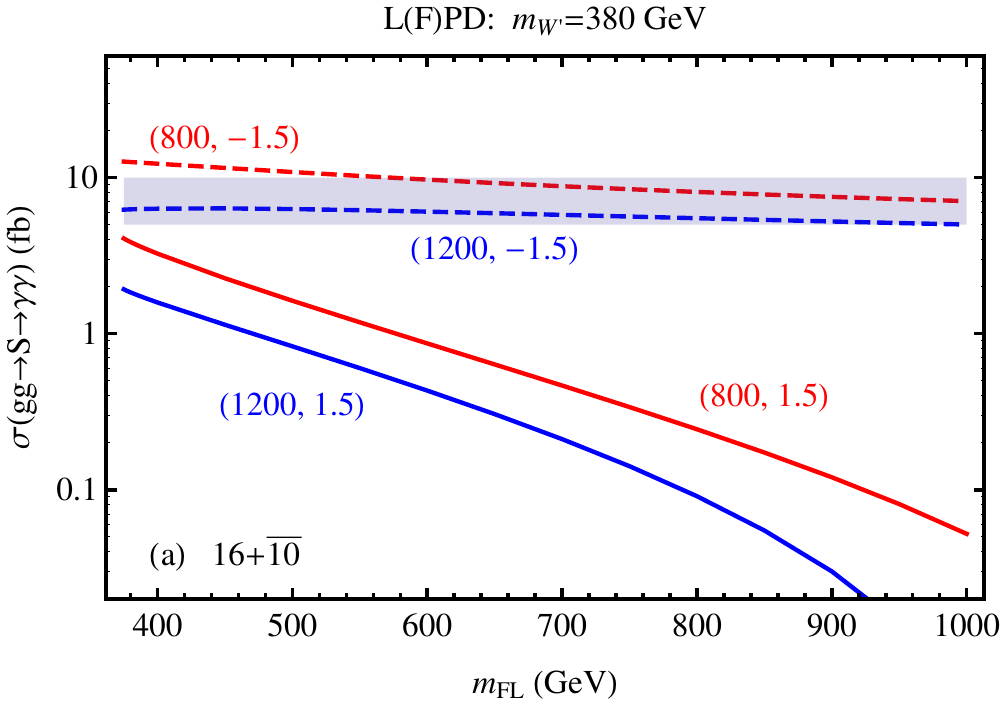}\quad\quad
\includegraphics[width=7.3cm]{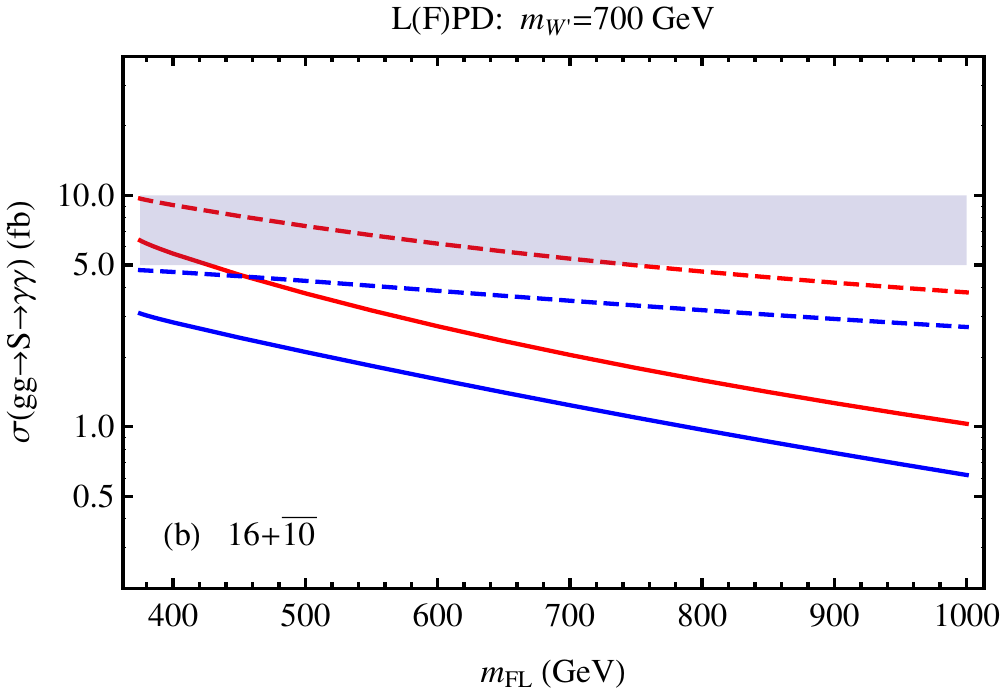}
\caption{$\sigma(gg\to S\to\gamma\gamma)$ as function of vectorlike lepton mass $m_{FL}$ for L(F)PD with (a) $m_{W'}=380\,$GeV, (b) $m_{W'}=700\,$GeV. We assume three family of $16+\overline{10}$. The color lines denote different combination of $(m_{FD}, y_\Delta)$ (mass in unit of GeV). The values are the same in two panels.}
\label{fig:Xsecaaf}
\end{figure*}

The second possibility is to have new colored vectorlike fermions below TeV scale.
Since the triplet scalar in L(F)PT models would not couple to colored or charged fermions as mentioned in Sec.\ref{sec:UV}, light vector-like fermions are irrelevant to diphoton signal. So we focus on the L(F)PD models here. As a remnant of $16_H$, the doublet $\Delta$ couples to vectorlike charged lepton and down-type quark in $16_F, 10_F$. As shown in (\ref{eq:VLFdelta}), they have the same yukawa coupling $y_\Delta$. But we could assume different mass $m_{FL}$, $m_{FD}$ for new leptons and quarks. Fig.\ref{fig:Xsecaaf} presents $\sigma(gg\to S\to \gamma\gamma)$ as function of $m_{FL}$ for three families of $16 \oplus \overline{10}$, assuming $m_{Z'}=3\,$TeV and $\tan\beta=0.01$. In each panel, the grey band shows the observed signal. The color lines denote different combination of $(m_{FD}, y_\Delta)$ (mass in unit of GeV), and the solid, dash lines for positive, negative $y_\Delta$ respectively. The vectorlike charged leptons interfere constructively with $W'$ in the loop when $y_\Delta<0$. As the diphoton branching ratio is large in this region, the signal rate is mainly determined by cross section of gluon-fusion production. This implies $m_{FD}\sim1\,$TeV, while the dependence on $m_{FL}$ is mild. With increasing $m_{W'}$, the interference becomes less significant and the signal rate drops faster with increasing $m_{FL}$.

There are LHC searches of vectorlike charged lepton and quark. The constraints on the fermion mass depend on the dominant decay modes. For vectorlike down-type quark, the lower bound is roughly $m_{FD}\gtrsim 700$\,GeV if it decays into $Hb$ or $Wt$, $Wq$ ($q$ is light quark). The constraint is pretty weak if $Hq$ is the dominant decay modes~\cite{Aad:2015kqa, Aad:2015tba}. The vectorlike charged leptons that contribute to diphoton loop belong to the $SU(2)_L$ doublet. They are mainly pair-produced in Drell-Yan processes. The lower bound is $m_{FL}\gtrsim 450\,(270)\,$GeV if it decays into $e/\mu$ ($\tau$)~\cite{Falkowski:2013jya}. Therefore, to accommodate diphoton excess in this model, among $W'$, new charged lepton and down-type quark, at least one cannot be far above the current bounds.

If the diphoton signature is confirmed, there should be matching signals on 750 GeV resonance in other channels at the Run 2 LHC with higher luminosity.
If the production and decays of the 750 GeV scalar are through the triangle loop,
the diphoton signal will be associated with $\gamma Z$, $ZZ$ and di-jet final states.
These additional channels provide cross-check on the diphoton signature.
However, in the above channels usually the new particles except the 750 GeV neutral scalar appear inside the loops of the diphoton process.
So it is hard to identify specific new physics models.

Direct productions of new physics particles provide us
a straightforward way to falsify various new physics models.
In the $G221$ models,
the direct production of new light $W'$ boson would be smoking gun signature of the LP and FP models.
For the LP model,
the following process
\bea
	p p \to W'  \to W Z
\eea
provide us signatures of the light $W'$ at the Run 2 LHC.
Although the $WZ$ channel has been searched at the LHC~\cite{Aad:2015ipg},
due to very small branching ratio of the $W' \to WZ$ (the dijet decay dominant),
the $WZ$ rate is expected to be at fb level.
We expect that the LHC Run 2 data could start to probe the interested parameter region.
In the FP model the $W'$ is very hard to probe directly as it doesn't couple to SM fermions.
The possible discovery channel could be
\bea
	p p \to \gamma^*/Z^* \to W'^+ W'^- \to W^+ Z W^- Z,
\eea
where $W$ and $Z$ could decay leptonic or hadronic.
In this model, the $W'$ decays dominantly to $WZ$ channel.
So the signal rate would not be suppressed by the small $W'WZ$ coupling.
Furthermore, this is a very clean channel with multi-lepton signature.
Hence we expect to observe the signal at the Run 2 LHC.
Since the 750 GeV scalar is the neutral component of the right-handed multiplet,
we may expect to see its charged counterpart in the same multiplet.
The doubly charged Higgs searches at the LHC Run 2 will be able to probe the favored parameter region in near future.


\section{Conclusion}
\label{sec:concl}

We investigated the possibility that the neutral component of the right-handed triplet or doublet $\Delta$ is identified as the 750 GeV scalar resonance $S$ to explain the diphoton excess observed at the LHC.
Based on the $SU(2) \times SU(2) \times U(1)$ effective theory framework,
we studied the breaking pattern I and new particle spectrum: $W'$, $Z'$ and scalar multiplet $\Delta$ , etc.
Although the current LHC searches put very tight constraints on these new particles, it is still possible to have  light $W'$ boson and/or $\delta^{++}$ in the LP and FP models.
The $SU(2) \times SU(2) \times U(1)$ framework could be easily incorporated into the Pati-Salam model and the $SO(10)$ GUT models.
In context of these UV models, there could be light colored scalars or light vectorlike fermions below TeV scale.

We found there are destructive interference between $W'$ and $\delta^{++}$ in the loop-induced diphoton process. Generally the $W'$ gives the dominant contribution, and the $\delta^{++}$ is important only if its mass is close to $m_S/2$.
Gluon fusion production of $S$ is initiated by loops of new colored particles. We studied the diphoton signal for both triplet or doublet $\Delta$ cases with two examples of colored scalars from $SO(10)$ GUT models, i.e. $SU(3)_c$ triple and octet. For colored vectorlike fermions, we focused on the doublet case with  new vectorlike fermions in $16+\overline{10}$ representation. For both scalars and vectorlike fermions, a light $W'$ is preferred if there is a considerable interference between $W'$ and new particles in diphoton loop.
To get a large enough gluon fusion production cross section, we need light (or a large numbers of heavy) colored scalars.  The color triplet of this low mass might be excluded by leptoquark search, but the color octet could still be light and thus provide us large enough diphoton signal. The vectorlike fermions are more strongly constrained. Assuming three families of $16+\overline{10}$, we can accommodate diphoton signal if among the $W'$, new charged lepton and down-type quark at least one has mass close to the current constraints.
If the diphoton signature is confirmed,
these benchmark models deserve a more careful study, including the particle spectrum, couplings, and various collider phenomenologies.

To accommodate the diphoton signature  in the LP or FP $G221$ framework, a light $W'$ boson would be a typical feature.
The direct searches of the light $W'$ is able to verify the $G221$ explanation.
The smoking gun signature of the LP and FP $W'$ is multi-gauge boson production with subsequent multilepton decays.
We expect the Run 2 LHC data provides us interesting signature associated with the diphoton signature.

\begin{acknowledgments}
JHY was supported in part by DOE Grant DE-SC0011095. J.R. is supported in part by the International Postdoctoral Exchange Fellowship Program of China. JHY would like to thank Tathagata Ghosh and Kuver Sinha for valuable discussions on the scalar sectors.
\end{acknowledgments}

\appendix

\section{Decay widths}
\label{app:decayw}

We parameterize the interactions of $S$ with charged vector boson and charged (and colored) scalars as following,
\begin{eqnarray}
\mathcal{L}\supset \lambda_{SV}SV^c_\mu V^{\mu}-\lambda_{S\chi_i}S \chi^c_i\chi_i-\lambda_{Sf_j}S\bar{f}_jf_j
\end{eqnarray}
The decay widths to gluon and photon are,
\begin{eqnarray}
  \Gamma_{S \rightarrow gg}
& = & \frac {\alpha_s^2}{128 \pi^3} m_S^3
\left| \sum_j C^j_R \frac{2\lambda_{Sf_j}}{m_{f_j}} A_{1/2}(\tau_{f_j}) +
  \sum_i C^i_R \frac{\lambda_{S\chi_i} }{m_{\chi_i}^2} A_0(\tau_{\chi_i})
\right|^2,
 \nonumber\\
  \Gamma_{S \rightarrow \gamma \gamma}
& = &
  \frac {\alpha^2}{1024 \pi^3} m_S^3
\left|
  \frac{\lambda_{SV}Q^2_V}{m_V^2} A_1(\tau_V) + \sum_j d^j_R \frac{2\lambda_{Sf_j} Q^2_{f_j}}{m_{f_j}} A_{1/2}(\tau_{f_j}) + \sum_i d^i_R \frac{\lambda_{S\chi_i} Q^2_{\chi_i}}{m_{\chi_i}^2} A_0(\tau_{\chi_i})
\right|^2,\nonumber\\
\end{eqnarray}
where $\tau_i \equiv 4 m^2_i / m^2_S$, $d_R^i$ is dimension of $SU(3)$ representation and $\textrm{Tr}(R_i^aR_i^b)=C^i_R\delta^{ab}$. The loop factors $A_i$ are defined as,
\begin{eqnarray}
A_1(\tau)&\equiv& 2 + 3 \tau [1 + (2 - \tau) f(\tau)],\quad\nonumber\\
A_{1/2}(\tau)&\equiv& -2\tau [1 + (1 - \tau) f(\tau)],\quad\nonumber\\
A_0(\tau)&\equiv& \tau [1 - \tau f(\tau)]
\end{eqnarray}
where
\begin{equation}
  f(\tau)
\equiv
\left\{\begin{array}{l}
  \left[ \sin^{-1} \left( \sqrt{1/\tau} \right) \right]^2 \,, \mbox{if } \tau \geq 1 \,, \\
- \frac 1 4 \left[ \ln \left( \eta_+ / \eta_- \right) - i \pi \right]^2 \,, \mbox{if } \tau < 1 \,,
\end{array}\right.
\end{equation}
with $\eta_\pm\equiv 1\pm \sqrt{1-\tau}$.

\end{document}